\documentclass[preprint,showkeys,aps,prb]{revtex4}
\usepackage{amsmath}
\usepackage[dvips]{graphicx}

\begin{document}

\title{
Time-Irreversibility is Hidden Within Newtonian Mechanics
}

\author{
William Graham Hoover and Carol Griswold Hoover \\
Ruby Valley Research Institute                  \\
Highway Contract 60 Box 601                     \\
Ruby Valley Nevada  89833 USA                   \\
}

\date{\today}


\vspace{0.1cm}

\begin{abstract}
We develop a bit-reversible implementation of Milne's Fourth-order Predictor algorithm so as to
generate precisely time-reversible simulations of irreversible processes.  We apply our algorithm
to the collision of two zero-temperature Morse-potential balls, which collide to form a warm
liquid oscillating drop.  The oscillations are driven by surface tension and damped by the
viscosities. We characterize the ``important'' Lyapunov-unstable particles during
the collision and equilibration phases in both time directions to demonstrate the utility of the
Milne algorithm in exposing ``Time's Arrow''.

\end{abstract}

\maketitle

\section{Introduction}
William Edmund Milne described half a dozen algorithms for solving ordinary differential equations in his 1949
book {\it Numerical Calculus}\cite{b1}, reprinted by Princeton University Press in 2015.  We were surprised to
find that two of his algorithms for first-order equations (an explicit algorithm on page 132, and an implicit
predictor-corrector algorithm on page 135) exhibit even-odd instabilities for the relatively undemanding
chaotic solutions of the Nos\'e-Hoover oscillator problem\cite{b2} with unit mass, force constant, temperature,
and Boltzmann's constant and a timestep $dt = 0.001$ :
$$
[ \ \dot q = p \ ; \ \dot p = -q - \zeta p \ ; \ \dot \zeta = p^2 - 1 \ ] \ ;
\ {\rm initially} \ (q,p,\zeta) = (2.4,0,0) \ .
$$
By contrast, the predictor stage of Milne's fourth-order (in $dt$) predictor-corrector algorithm (page 140)
for second-order Newtonian motion equations is more useful. It can be adapted to generalize Levesque and
Verlet's second-order-accurate bit-reversible algorithm\cite{b3} to a fourth-order bit-reversible integrator
with global trajectory errors similar to the fourth-order errors of the classic Runge-Kutta integrator. This
development provides bit-reversible trajectories capable of being integrated reversibly as far as one likes,
backward and forward in time, eliminating the need to store them when time-reversed analyses are desired. The
local error of Milne's page-140 predictor is $(17dt^6/240)$. Here we apply this explicit algorithm to an
irreversible process, the inelastic collision of two similar balls each composed of 100 particles.  The
Morse potential is satisfactory for this process, making it possible to study in detail the dynamical
instabilities associated with irreversible flows in systems obeying Newtonian (or Hamiltonian) classical
mechanics and illustrating Loschmidt's Reversal and Zerm\'elo's Recurrence Paradoxes.

We begin by reviewing the disparity between Newtonian time reversibility and the irreversibility of the Second
Law of Thermodynamics.  This motivates our interest in bit-reversible integrators, extending the work of
Levesque and Verlet to the more-nearly-accurate alogorithm developed by Milne. A bit-reversible analysis of the
Lyapunov instability of the colliding balls, both forward and backward in time, illustrates the nonphysical
nature of the reversed trajectory.  We conclude with a Summary and Recommendation.

\section{Newtonian Models Describe Some Irreversible Processes}

The disparity between time-reversible atomistic mechanics and real-life irreversibility has been a
scientific discussion subject ever since Boltzmann's H Theorem, Loschmidt's Reversal Paradox,
Maxwell's Demon, and the Poincar\'e-Zerm\'elo Recurrence Paradox.  Versions of ``Humpty Dumpty Had a
Great Fall'' preceded all these worthies -- ``All the King's horses and all the King's men couldn't put
Humpty together again''.

The theoretical time-reversible nature of dynamics doesn't often carry over to the numerical algorithms
used in simulation\cite{b3}. Computational rounding errors, soon amplified by Lyapunov instability to grow
exponentially in time, characterize chaotic systems, with the springy pendulum\cite{b4} and the periodic
Lorentz gas\cite{b5} providing examples with two degrees of freedom, the minimum for chaos. Alexander
Lyapunov analyzed the instability of mechanical flows in terms of the rates of divergence of the distance
between two nearby trajectories in phase space.  The time-averaged rate defines the largest Lyapunov
exponent, $\lambda_1 \equiv \langle \ \lambda_1(t) \ \rangle$, where $\lambda_1(t) $ is a ``local'' or
``instantaneous'' rate .  Additional exponents describe the growth or decay rates of two-dimensional,
three-dimensional, ... , N-dimensional volumes in the $N$-dimensional space required to describe the
flow\cite{b6,b7}.

Hamiltonian mechanics cannot describe nonequilibrium steady states\cite{b8,b9}.  They are intrinsically
irreversible.  By contrast, Levesque and Verlet pointed out that an integer version of the Leapfrog
algorithm can precisely reverse dynamics in just the way visualized by Loschmidt in his Reversibility
objection to Boltzmann's H Theorem\cite{b2}.  The bit-reversible integer algorithm has the form :
$$
\{ \ q_{t+dt} - 2q_t + q_{t-dt} = a_tdt^2 \ \} \ [ \ {\rm Levesque-Verlet} \ ] \ .
$$
Here all four terms are integers. Even given the initial conditions (two adjacent sets of coordinates) the
trajectories produced are not unique.  In addition to sensitive dependence on the timestep $dt$ W. Nadler,
H. H. Diebner, and O. E. R\"ossler\cite{b10} pointed out that one can ``round'' the acceleration term $adt^2$
(either ``up'' or ``down'') rather than simply changing it to an integer closer to, or farther from, zero. 
Whether or not the three variants can have ``interesting'' consequences is not known.

Integrating the fourth-order ``local'' single-step  error term $\frac{1}{12}\stackrel{....}{q}dt^4$ twice with
respect to time gives ``global'' errors that are second-order in $dt$. Representing each of the four terms as
a (``large'', for instance 15-digit or 30-digit) integer makes it possible to extend a finite-difference
caricature of a Newtonian trajectory forward or backward in time, arbitrarily far and with perfect reversibility.

This computational reversibility makes it possible to explore both classic objections to the use of Hamiltonian
mechanics as an explanation of irreversible behavior.  Loschmidt seized on the Reversibility of Hamiltonian
mechanics. Any ``irreversible process'' is obviously impossible with a ``reversible'' dynamics.  The difficulty
in imagining an irreversible process governed by Hamiltonian mechanics extends also to the case in which
Hamiltonian mechanics includes heat reservoirs imposed by Lagrange multipliers or by Nos\'e's Hamiltonian\cite{b8}.
Hamiltonian or Lagrangian approaches to nonequilibrium simulations fail to illustrate the Second Law and in fact
only illustrate the failure of Lagrangian or Hamiltonian thermostats to generate nonequilibrium steady states.
Gaussian and Nos\'e-Hoover dynamics, on the other hand, do allow for steady heat transfer, the fundamental mechanism
underlying the Second Law and allowing change in the comoving phase volume in a way completely consistent with
Gibbs' phase-volume definition of the thermodynamic entropy\cite{b11}.

Zerm\'elo pointed out that a confined, but otherwise isolated, purely-Hamiltonian system will in principle approach its
initial state arbitrarily closely. Though this ``recurrence'' is certainly correct mathematically, algorithms for
digital computers typically generate instead a transient trajectory ultimately leading to divergence, or a periodic
orbit, or a fixed point.  Evidently the Levesque-Verlet computer algorithms {\it are} reversible.  As a consequence
Levesque and Verlet's multidigit approximations to a Newtonian trajectory all exhibit the perfect Reversibility that
was Loschmidt's target as well as the near-perfect Recurrence objection of Zerm\'elo, based on Poincar\'e
recurrence. Both objections appear to contradict the oneway nature (``entropy must increase'') of the Second Law
of Thermodynamics.

In practice we have used the Levesque-Verlet algorithm to study the lack of symmetry in the two Lyapunov vectors
$\{ \ (\delta_q,\delta_p) \ \}$, one generated backward and the other forward, associated with the largest Lyapunov
exponents, $\lambda_1(\pm t)$, with the two divergence rates followed in the two time directions. Followed forward
in time the particles contributing the most to that exponent are exactly those intuitively expected from macroscopic
phenomenology --- those prominent in entropy production.\cite{b12,b13}.  But studying Lyapunov's divergence rates near this
same trajectory but in the opposite time direction reveals that wholly different particles become ``important'', making
above-average contributions to $\lambda_1(t)$. The reversed trajectories are characteristically
{\it inconsistent} with macroscopic phenomenology.  This disparity shows that Lyapunov's analysis singles
out the ``past'', with a qualitative difference between the Lyapunov instabilities seen forward and backward in time.
In fact if one calculates the phenomenological flow stress or transport coefficients in a reversed trajectory he finds
that they have the wrong sign in the reversed flow.  In that reversed flow
heat moves preponderantly from cold to hot and the stress, which is independent of the time direction, appears to respond
to the local strain rate with a negative viscosity coefficient or a negative yield stress.\\

This same precise reversibility of the Levesque-Verlet algorithm can also be achieved\cite{b13} with the
higher-order page-140 bit-reversible Predictor algorithm from Milne's book :
$$
\{ \ q_{t+2dt} - q_{t+dt} - q_{t-dt} + q_{t-2dt} \ \} = (dt^2/4)[ \ 5a_{t+dt} + 2a_t + 5a_{t-dt} \ ]
\ [ \ {\rm Milne \ Predictor} \ ] \ .
$$

The local error this algorithm incurs is sixth-order in $dt$ so that two integrations with respect to time
give a fourth-order global error, the same order of accuracy as the classic Runge-Kutta RK4 algorithm but
with the advantage of precise time reversibility.  Using Milne's predictor algorithm the harmonic oscillator
problem has an analytic solution\cite{b13}.  {\bf Figure 1} compares the phase-shift errors for the Predictor
and RK4\cite{b14} algorithms for the oscillator, establishing that both integration methods are fourth-order.  The
Figure also shows that the Milne predictor error is about 40\% greater than the Runge-Kutta error, a negligible
difference for our problems.

\begin{figure}
\center
\includegraphics[width=2.2in,angle=-90,bb=48 22 567 750]{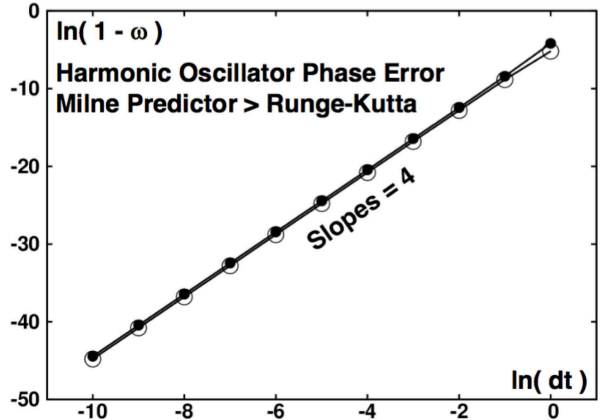}
\caption{
The fourth-order Runge-Kutta (open circles) and Milne Predictor (filled circles) integrators have similar accuracies
when applied to the harmonic oscillator problem $\ddot q = -q \rightarrow q = \cos(\omega t)$ with $\omega = 1$.
Here we see that the dominant phase-shift error, $1-\omega$, varies as the fourth power of the timestep $dt$.
Both these problems have analytic solutions useful for developing new software\cite{b13,b14}.
}
\end{figure}

For large systems Milne's method, with only one force evaluation per timestep, is more ``efficient'', though
the required storage is about twice as great . The combination of fourth-order Runge-Kutta with fourth-order
Milne provides the tools for highly-accurate solutions of the many-body problem that can readily be followed,
forward or backward, without difficulty.  A sixth-order velocity definition from page 99 of Milne's book,
$$
\dot q_t \equiv {\textstyle (\frac{3}{ 2}) }\frac{q_{t+ dt}-q_{t- dt}}{2dt}
               -{\textstyle (\frac{3}{ 5}) }\frac{q_{t+2dt}-q_{t-2dt}}{4dt}
               +{\textstyle (\frac{1}{10}) }\frac{q_{t+3dt}-q_{t-3dt}}{6dt} \ ,
$$
contributes accurate values of the kinetic temperature, stress tensor, and heat flux vector when used in
conjunction with the two fourth-order integrators.  We will demonstrate the application of these ideas using
the simple Morse pair potential, the difference of two exponential functions. Let us first explain our reasons
for this choice.

\section{Choosing a Morse-Potential System for Liquid Drops}

In joint work with Karl Travis and Amanda Hass (University of Sheffield) we had planned liquid-drop simulations using
the ``84'' pair potential, $\phi(r<\sqrt{2}) = (2-r^2)^8 - 2(2-r^2)^4$, chosen for its simplicity, smoothness,
and short range.  We soon discovered the lack of a liquid phase in our simulations\cite{b15}.  Long ago
in 1993 Daan Frenkel, along with four of his colleagues\cite{b16}, had done some work noting the lack of a
liquid phase for C$_{60}$.  In the midst of our liquid-drop work Karl forwarded the Frenkel paper along to
us, ultimately giving rise to this Festschrift contribution.  The missing liquid phase is caused by the
relatively short range of the potential, causing surface particles' binding energies to be so much smaller
than the bulk binding energy that a heated solid sublimes rather than melting. Thus our first concern was
finding a simple longer-range potential that would encourage the liquid phase. Our own interest in
confronting time-reversibility with the Second Law was temporarily put on hold while seeking a suitable pair
potential. 

Because the Lennard-Jones and 84 potentials with which we started out are evidently unable to support a
liquid-gas interface for a small number of particles we turned next to searching for other forcelaws in
which the binding energy of surface particles is strong enough to stabilize a fluid phase. The Morse
potential, with its separation and depth at minimum setting the length and energy scales, still provides
a variety of attractive wells depending on the stiffness parameter $\alpha$ :
$$
\phi(\alpha) =   e^{-2\alpha(r-1)} - 2e^{-\alpha(r-1)} \ .
$$
Sample Morse potentials are compared with the Lennard-Jones potential in  {\bf Figure 2}. For convenience
we choose to set both the particle mass and Boltzmann's constant equal to unity in this work, along with
the distance and energy scales from the potential.

\begin{figure}
\center
\includegraphics[width=2.in,angle=-90,bb=181 12 429 782]{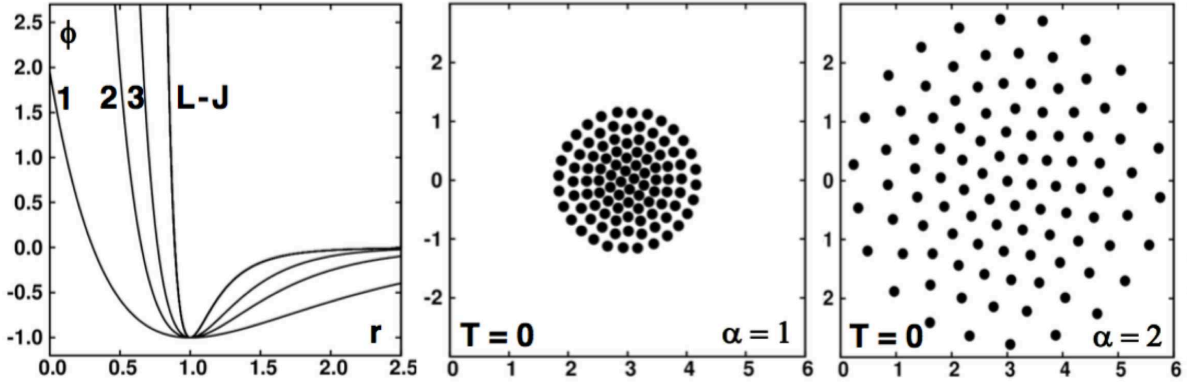}
\caption{
The Lennard-Jones pair potential is compared with Morse potentials with $\alpha =$ 1 and 2 and 3.
Our time-reversible colliding-ball simulations are based on the choice $\alpha=2$, with a binding
energy of order 8 per particle providing the larger nearly-circular zero-temperature structure shown 
at the right. The smaller ball has $\alpha = 1$ and a much higher density due to the wide potential
bowl.
}
\end{figure}

Choosing $\alpha$ equal to two for further investigation we carried out a series of Nos\'e-Hoover
runs in which cold hundred-particle balls were slowly heated to a kinetic temperature
of $T_\infty = \langle \ (K/N) \ \rangle$ :
$$
\{ \ \ddot q = a_\alpha - \zeta \dot q \ \} \ ; \
\dot \zeta = \sum [ \ \dot q^2 - (t/t_{\rm max})T_\infty \ ] \ ; \ t_{\rm max} = 1000 \ .
$$
Here the $\{ \ a_\alpha \ \}$ are the conventional Newtonian accelerations. The Nos\'e-Hoover friction
coefficient $\zeta$ controls the progress of the temperature to $T_\infty$.  {\bf Figure 3} shows
that at a final temperature of 0.05 the ball has deformed just a little, mostly taking on a triangular
lattice structure with a few lattice defects.  The low-temperature deformation of that solid structure
occurs through the occasional ``hexatic'' sliding of rows of particles.  At the higher temperature of
0.25 the ``ball'' has melted to become a ``drop''. In the drop ordinary diffusion dominates the
deformation.  In both these snapshots the ball and drop rotate slowly while the center-of-mass is
motionless. With a potential able to stabilize the liquid phase let us proceed to a solid-to-liquid
application of the Morse potential simulating an irreversible process.

\begin{figure}
\center
\includegraphics[width=2.in,angle=-90,bb=112 15 488 781]{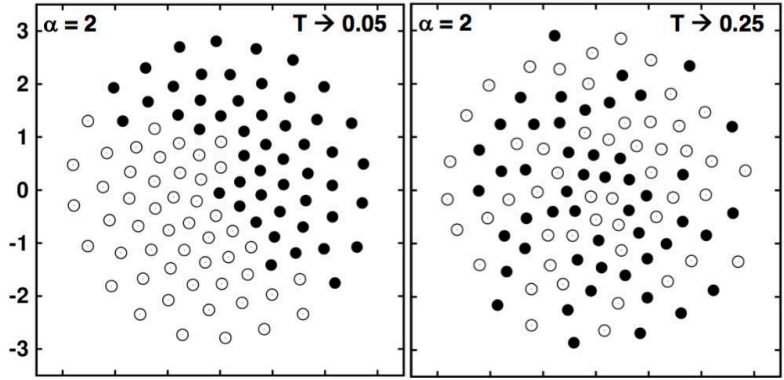}
\caption{
  Cold hundred-particle $T=0$ balls were gradually heated (over a time of 1000) to temperatures of 0.05
(left) and 0.25(right).  The lower temperature resulted in a solid undergoing occcasional sliding-row
deformations. The higher temperature ball is a typical liquid. The two plotting symbol types indicate
particles originally on opposite sides of the balls' centers.
}
\end{figure}

\section{Choosing an ``Irreversible'' Hamiltonian Problem}

\begin{figure}
\center
\includegraphics[width=3.0in,angle=0,bb= 8 92 597 704]{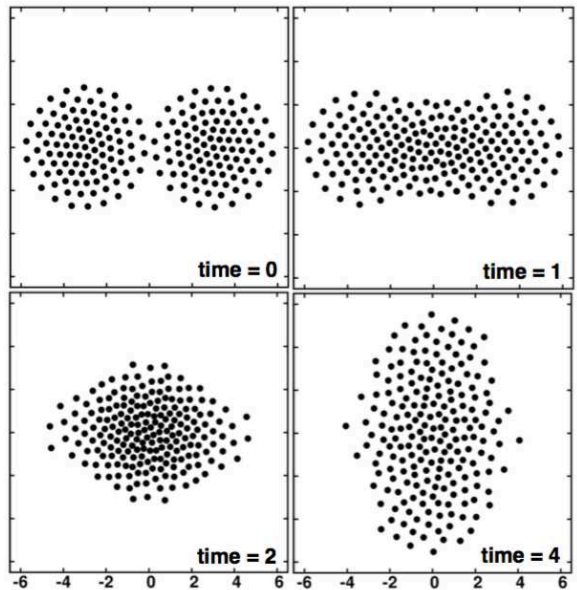}
\caption{
The initial condition of two zero-temperature Morse balls with $\alpha = 2$ is at upper left.  Three snapshots
show the extremes of $x$ and $y$ displacement early in the equilibration process.
}
\end{figure}

To the Question ``What is the simplest localized irreversible process we can simulate with Hamiltonian 
mechanics?'' the Answer is plain, ``A head-on two-body collision of two similar quiescent bodies.''  For
simplicity we choose exactly similar bodies with vanishing center-of-mass and angular momentum, converting
all of the initial excess surface and interaction energy to heat.  There are no boundary conditions to
implement and we expect the system to equilibrate after a few sound traversal times. A smooth force law
which stabilizes the final equilibrium state, a fluid drop, is the goal. Smoothness enhances energy
conservation and allows for larger timesteps. Typical Runge-Kutta or Milne simulations with 100,000 timesteps
$dt = 0.001$ conserve the energy to 9-figure accuracy.

Our 2015 bit-reversible Levesque-Verlet simulations\cite{b17} used relatively intricate forces including
a smooth-particle density-dependent attractive force together with a pairwise-additive repulsion.  The
Morse pair potential, with a large readily accessible liquid range is perfect for irreversibility studies.
In {\bf Figure 4} we show three snapshots from the equilibration of two zero-temperature balls initially
close together, shown at the upper left. {\bf Figure 5} shows the variation of the kinetic energy with
time, driven by surface tension and damped by shear and bulk viscosities, as discussed by Nugent and
Posch\cite{b18}.  We will analyze similar configurations' dynamical instabilities both forward and
backward (reversed) in time in what follows.

A reproducible protocol for creating the two interacting nearly circular balls shown at the top left of
{\bf Figure 4} begins by relaxing a $10 \times 10$ square of 100 particles with nearest-neighbor spacing
of unity and centered on the origin.  For this relaxation we choose the Morse parameter $\alpha= 1.5$ with
a friction coefficient of unity giving a damping force $-p$.  The roughly circular zero-temperature
fully-relaxed static configuration (not shown here) is then allowed to expand and relax again after
changing $\alpha$ from 1.5 to 2.0. Again the relaxation forces are $-p$. Moving that relaxed structure to
the right a distance 3 or more, and then constructing an inversion-symmetric twin with inverted coordinates
and velocities :
$$
\{ \ x(i+100) = -x(i) \ ; \ y(i+100) = -y(i) \ ; \ p_x(i+100) = -p_x(i) \ ; \ p_y(i+100) = -p_y(i) \ \} 
$$
gives a suitable initial condition for our irreversible two-ball collision process.  The long-range nature
of the Morse potential causes the two separated balls to ``collide''.  The initial condition, with a 
center-to-center disance of 6, and three later snapshots of the system are shown in {\bf Figure 4}.

Unlike a typical scattering problem we start here with the two balls motionless, a ``turning point'' for all
of the atoms.  For this special choice the past and future are exactly alike! A stochastic analog is the
usual initial condition in the Ehrenfests' dog-flea model, with all the fleas on just one of the dogs, an
unlikely state as the fleas jump stochastically, one at a time.  From the mathematical standpoint our initial
condition has perfect inversion symmetry.  From the computational standpoint this symmetry could soon disappear.
When sequences of mathematical operations are performed in different orders the roundoff errors can differ.
This difference can occur in the two-ball problem and then grows exponentially (due to Lyapunov instability)
unless the computation is symmetrized. Inversion symmetry at every timestep can be imposed by averaging or by
choosing one of the balls to impose its inverted configuration on the other one. We choose to apply the
inversion equations just given at the conclusion of each timestep.

\begin{figure}
\center
\includegraphics[width=2in,angle=-90,bb=48 22 567 750]{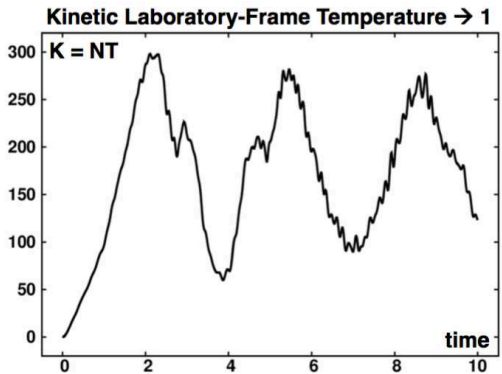}
\caption{
The early oscillations of temperature converge to a final temperature of unity.
}
\end{figure}

The corresponding laboratory-frame kinetic temperature is shown in {\bf Figure 5}.  Later the kinetic energy
fluctuates about 200, so that $T = 1$. The zero-temperature single-ball initial condition shown at the
right in {\bf Figure 2}, leading to Nos\'e-Hoover snapshots at temperatures in the range [ 0.05 to 1.00 ] indicated
that a heated $\alpha = 2$ ball becomes a stable fluid or a hexatic solid, with no sublimation throughout
that range, as shown in {\bf Figure 3}. Particles originally on the right and left of the initial ball are
distinguished at $T = 0.05$ and 0.25 in {\bf Figure 3}.  This shows that the 200-particle compound drop is indeed
liquid. In {\bf Figure 4} snapshots at times of 1, 2, and 4 show the formation of a 200 particle liquid drop with
total energy $-1788 \simeq -9N$. At longer times the mean kinetic energy fluctuates about $K = N$. We found
no sublimation in extending the isoenergetic simulation of {\bf Figures 4 and 5} for a million timesteps to a time
of 1000 with $dt = 0.001$. Let us consider the details of monitoring the Lyapunov instability of the collision
process with Milne's integrator.

\section{Bit-Reversible Simulations of the Colliding Balls}

In developing our present simulations with relatively complicated motion and velocity algorithms, program
simplicity was our paramount goal. Combining Milne's Predictor with the sixth-order velocity algorithm
described in Section II requires keeping seven successive integer coordinates, covering the time range
from $t-5dt$ to $t +dt$ . Three of these floating-point values of the coordinates are required for the
accelerations.  The local Lyapunov exponent, one from a forward analysis and one from its backward twin, are
both determined by constraining the separation between a classical conservative reference trajectory and a
satellite trajectory.  The satellite is constrained to remain at a fixed distance from the reference.  Milne's
Predictor algorithm used for the ``reference trajectory'' is time-reversible to the very last bit. The RK4
algorithm used for the satellite, which is rescaled to distance $\delta = 0.00001$ from the reference at every
timestep, becomes equivalent to a Lagrange-multiplier constraint as $dt \rightarrow 0$. The Milne trajectory can be
reversed by simply inverting the time order of seven successive configurations. The direction chosen by the
reference-to-satellite vector converges to machine accuracy at a time of order 10, with the corresponding
Lyapunov exponent visually reversible for a time of order 2.

Although the time-reversal programming seems daunting it can be accomplished simply. Switching the
order of the last reference trajectory coordinates along with their finite-difference velocities and
reversing the ordering of the Runge-Kutta satellite variables over that same interval, of length
$6dt= 0.006$, is all that is required. The algorithmic errors are both of order $dt^4$ and quite
negligible over such an interval. Using the roughly-circular balls formed with $\alpha = 2$,
gives a relaxed potential energy near -850. Placing two such crystallites on the $x$ axis
with a center-to-center spacing of 6 (see again {\bf Figure 4}) gives a total energy of -1788. Allowing these
crystallites to equilibrate without further friction gives an equilibrated kinetic energy close to 200,
corresponding to a kinetic temperature of unity.  A wider center-to-center spacing of 8 gives instead a total
energy of -1708 and an equilibrated kinetic energy of 250, while 10 gives -1705.  {\bf Figure 6} details the
history of the laboratory-frame kinetic energy, showing half a dozen oscillations of the warming drop prior
to equilibrium. This simulation began with a center-to-center spacing of 10.  Let us turn to a few details of
the simulation.

\begin{figure}
\center
\includegraphics[width=3.in,angle=-90,bb=31 17 578 781]{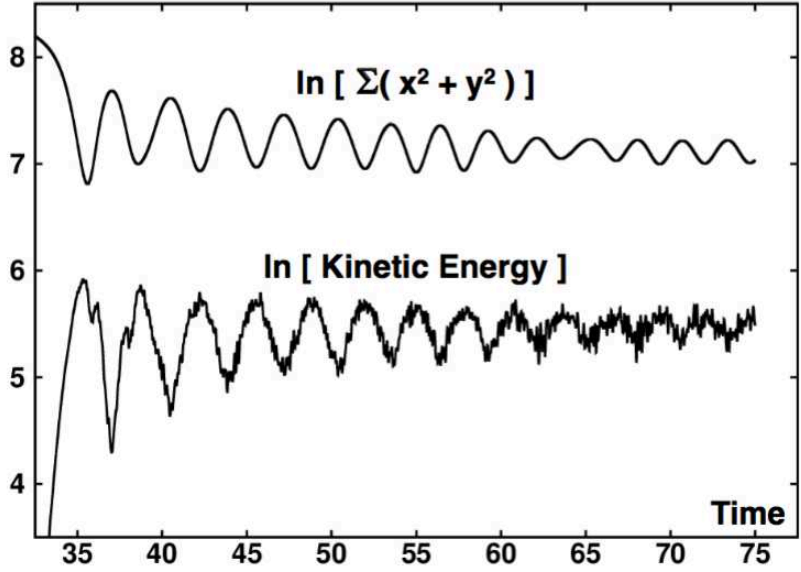}
\caption{
Time variation of the kinetic energy and $\sum r^2$. Each vibrational cycle corresponds to two maxima and
minima as the oscillating drop stretches alternately in the $x$ and $y$ directions.
}
\end{figure}

\section{Levesque-Verlet and Milne Bit-Reversible Algorithms}

The algorithms for a bit-reversible simulation, coupled with an analysis of the largest Lyapunov exponent
and its offset vector are most simply approached in two separate parts.  We begin with a description of
the bit-reversible programming for the harmonic oscillator and then describe the additional work required
for the Lyapunov analysis.

\subsection{Oscillator Solutions and Momentum Definitions}

It is simplest to develop bit-reversible algorithms by solving special cases of the harmonic oscillator
problem with known solutions. The simplest Levesque-Verlet oscillator problem has a timestep {\tt DT = 1},
a period of 6 rather than $2\pi$, and initial turning-point coordinates of {\tt QM = Q = 1}. The equation
of motion and its periodic solution are :
\begin{verbatim}
QP = 2*Q - QM - Q*DT*DT = Q - QM
\end{verbatim}
$$
\longrightarrow \{ \ Q \ \} = \{ \ 0,-1,-1,0,1,1,0,-1,-1,0,1,1,\dots \ \} \ .
$$

Similarly, the Milne oscillator with a timestep {\tt DT = 2} and a period of 8, as opposed to $2\pi$, can be
generated with initial conditions {\tt QMMM = 1; QMM = 0; QM = -1; Q = 0} with the motion equation and its
periodic solution :
\begin{verbatim}
QP = Q + QMM - QMMM - (5*Q + 2*QM + 5*QMM)*DT*DT/4
\end{verbatim}
\begin{center}
or
\end{center}
\begin{verbatim}
QP = -4*Q - 2*QM - 4*QMM - QMMM
\end{verbatim}
$$
\longrightarrow \{ \ Q \ \} = \{ \ 1,0,-1,0,1,0,-1,0,1,0,-1,0, \dots \ \} \ .
$$
For other timesteps the initial values of {\tt QMMM} \dots {\tt Q} can be generated by Runge-Kutta with a
negative timestep or from the analytic cosine solution.  For phase-space Lyapunov analyses momenta are also
required. These can be {\it defined} by second-, fourth-, or sixth-order centered differences :
$$
\dot x_t \equiv                            \frac{x_{t+ dt}-x_{t- dt}}{2dt} \ ;
$$
$$
\dot x_t \equiv {\textstyle (\frac{4}{ 3}) }\frac{x_{t+ dt}-x_{t- dt}}{2dt}
               -{\textstyle (\frac{1}{ 3}) }\frac{x_{t+2dt}-x_{t-2dt}}{4dt} \ ;
$$
$$
\dot x_t \equiv {\textstyle (\frac{3}{ 2}) }\frac{x_{t+ dt}-x_{t- dt}}{2dt}
               -{\textstyle (\frac{3}{ 5}) }\frac{x_{t+2dt}-x_{t-2dt}}{4dt}
               +{\textstyle (\frac{1}{10}) }\frac{x_{t+3dt}-x_{t-3dt}}{6dt} \ .
$$
In implementing the bit-reversible algorithms 15- or 30-digit integer values of the coordinates are desirable.
In FORTRAN77 this is accomplished by
\begin{verbatim}
INTEGER*8 IQMMM, IQMM, IQM, IQ, IQP
\end{verbatim}
\begin{center}
 or 
\end{center}
\begin{verbatim}
INTEGER*16 IQMMM, IQMM, IQM, IQ, IQP .
\end{verbatim}

\subsection{Calculation of the Largest Lyapunov Exponent}

Here we assume the availability of a phase-space reference trajectory, $\{ \ (q,p)_{t\leq 0} \ \}$, generated
by one of the bit-reversible algorithms just illustrated.  We also imagine that a satellite trajectory has been
generated with a piecewise-accurate Runge-Kutta integration with the result $\{ \ (\delta_q,\delta_p)_{t\leq 0} \ \}$
for all time values up through the current time, time zero.  We assume that the ``scaling step" has been done as well
so that the length of the offset vector is $\delta = 10^{-5}$.  Here follow the steps required to generate all of
these variables at time $dt$ :\\
\noindent
[ 1 ] Integrate Hamilton's motion equations for the satellite trajectory from $\{ \ (q+\delta_q,p+\delta_p)_0 \ \}$ to an
unconstrained $\{ \ (q+\delta_q,p+\delta_p)_{dt(new)} \ \}$ using fourth-order or fifth-order Runge-Kutta integration.\\
\noindent
[ 2 ] Determine the integer-based values of $\{ \ (q,p)_{dt} \ \}$ according to a bit-reversible integrator,
Levesque-Verlet for second-order accuracy and Milne for fourth-order.\\
\noindent
[ 3 ] Find the length of the new offset vector and the scale factor needed to return it to its original value $\delta = 0.00001$ .
$$
\delta_{dt}^{\rm new} \equiv \sqrt{\sum [ \ (q+\delta_q,p+\delta_p)_{dt} - (q,p)_0 \ ]^2 } \ ; \
\lambda(t) \equiv -\ln[ \ \delta_{dt}^{\rm new}/\delta \ ]/dt \ .
$$
\\
\noindent
[ 4 ] Determine the scaled offset vector with length $\delta = 10^{-5}$ giving the local Lyapunov exponent
$\lambda(dt)$ at time {\tt dt} and the scaled value of the offset vector $\{ \ (\delta_q,\delta_p) \ \} \ $.\\
\noindent
[ 5 ] Shift all ``new'' variables, $\{ \ {\tt IQP}, \ {\tt IPP} \ \}, \{ \ qp, \ pp, \ \delta_qp, \ \delta_pp \ \}$ to
``older'' locations. Here the final $p$ indicates the time, not momentum !

We believe that diligent implementation of these ideas will allow the reader to reproduce our results and
explore some of the fascinating problems arising in the simulation of conservative nonequilibrium flows. An
excellent aid to programming the Milne and Lyapunov algorithms is to apply both techniques to a simple
problem with an analytic solution. The harmonic oscillator suggests itself as such a problem.  We have seen
that the Levesque-Verlet and Milne algorithms both have analytic solutions for that problem.  In order to
include Lyapunov instability in the analysis it is only necessary to ``scale'' the oscillator, converting
its circular orbit to an ellipse\cite{b19}.

\subsection{The Scaled Oscillator, a Useful Warmup Problem}

An elliptical orbit, with width 4 and height 16, is a solution of the scaled oscillator Hamiltonian:
$$
{\cal H} = (1/2)[ \ (q/s)^2 + (ps)^2 \ ] = 8 \ ,
$$
where the scale factor $s$ is (1/2). With initial values $(q,p) = (2,0)$ the form of the solution is
$q = 2\cos(t) \ ; \ p =-8\sin(t) $ . The infinitesimal offset vector components, $(\delta_q,\delta_p)$
themselves obey the equations :
$$
\{ \ \dot q = (p/4) \ ; \ \dot p = 4q \ \} \longrightarrow
\{ \ \dot \delta_q = (\delta_p/4) - \lambda \delta_q \ ;
\ \dot \delta_p = -4\delta_q - \lambda \delta p \ \} \ ,
$$
with the solution shown in {\bf Figure 7}. The local largest Lyapunov exponent $\lambda_1(t)$ is equal
to the Lagrange  multiplier $\lambda$ which is chosen to satisfy the constraint $\delta_q^2 + \delta_p^2
\equiv \delta^2$. The solution $\lambda = -(15/4)\delta_q \delta_p/\delta^2$ is shown in {\bf Figure 7}.
With this problem as a guide a determined reader should be able to develop programs for Milne
integration for a reference trajectory coupled to Runge-Kutta integration for one or more
satellite trajectories.  On a multicore machine the entire Lyapunov spectrum for a manybody system
can be obtained in this way.

\begin{figure}
\center
\includegraphics[width=2.0in,angle=-90,bb=151 15 457 774]{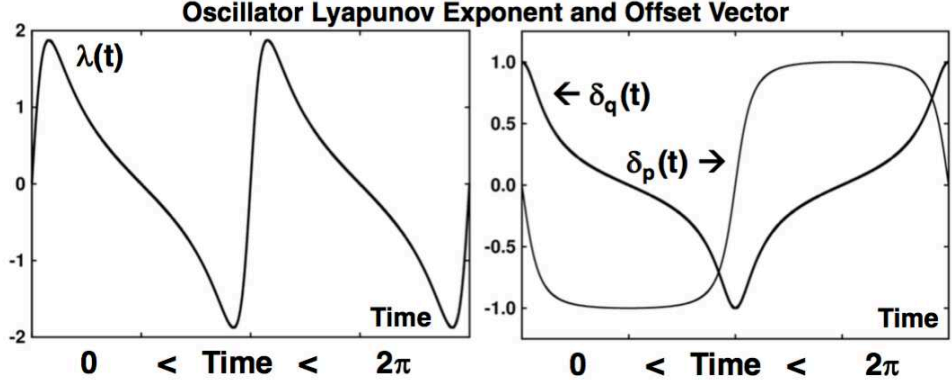}
\caption{
The local Lyapunov exponent, $\lambda_1(t) = -(15/4)\delta_q(t)\delta_p(t)/\delta^2$ (left) and offset
vector components $(\delta_q,\delta_p)$ (right) for the scaled-oscillator problem with the scale factor
$s=(1/2)$ .
}
\end{figure}

\section{Sample Results from Two-Ball Collision Problems}

Preliminary simulations show that the local Lyapunov exponent $\lambda_1(t)$ requires a time of order 2
to converge. Calculations with two quiescent balls with initial center-to-center distance of 6 as in
{\bf Figure 4} were analyzed first. The local Lyapunov exponent for the particular $t=2000dt=2$
configuration compared $\lambda_1(t=2)$ going forward and backward in time for intervals of 18-16, 10-8,
6-4, and 4-2. The perfect agreement of the Lyapunov exponents at the end time +2 of these simulations,
indicated that the relaxation time of the offset vectors is no larger than 2. Accordingly, in order to
allow time for the offset vectors to reach a converged orientation our remaining runs began with a
conservative initial spacing of 10 rather than 6.

{\bf Figure 6} showed the close correspondence between oscillations in the kinetic energy and the squared
amplitude of oscillation. The frequency of oscillation was related to surface tension by Rayleigh and
was confirmed by Nugent and Posch in analyzing their two-dimensional smooth-particle simulations\cite{b18}.
The local Lyapunov exponent requires the rescaling of the offset vector at each time step.  At
the reversal of the trajectory the exponent changes sign precisely so that $\lambda_f$ and $\lambda_b$ sum
exactly to zero until Lyapunov instability destroys their correlation.  After a further
time of about 2 the reversed exponent becomes independent of the reversal time.  Thus we can obtain accurate
values of the local Lyapunov exponent and its offset vector in both time directions by adding a few thousand
timesteps at each end of the run.  It is not necessary to store any intermediate configurations as the Milne
algorithm is precisely reversible.

The long-time-averaged Lyapunov exponent $\lambda_1 \equiv \langle \ \lambda_1(t) \ \rangle$ for the
equilibrated drop is about 5 so that the local exponent remembers a history of about
$e^{5t} = 10^{15}\rightarrow t = 7$. To avoid spatial asymmetry in the vectors corresponding to $\lambda_1(t)$
we also symmetrize the contributions of the balls to the offset vector $\delta_{s-r}$  When this is done
the important particles in the collision process turn out to be totally different in the two time directions.
After trajectory reversal a convergence time of 8 is more than sufficient for machine-accurate agreement of the
important particles in the reversed trajectory.  Our local Lyapunov analysis identitifies those particles
making above-average contributions to the offset
$$
\delta(t) = \sqrt{ \ \sum [ \ \delta_x^2 + \delta_y^2 + \delta_{p_x}^2 + \delta_{p_y}^2 \ ] } \ .
$$ 
Lyapunov analysis augments the local description by identifying the contributions of degrees of freedom to
dynamical instability.  Although analogous to temperature or energy Lyapunov instability contains the Arrow of
Time.  Let us consider a typical vibrational cycle to learn how the important particles are identified
in both time directions.

Initially all the particle velocities in both balls are zero. Plots of kinetic energy and $\sum (x^2 + y^2)$
correspond nicely as was shown in {\bf Figure 6}. Two maxima correspond to a drop oscillation, with elongation
of the drop first maximum in the $x$ direction and then the $y$. The collision begins at a time of about 30 and
was followed to times of 50 and 100 and back. These two choices agree precisely in our region of interest
from 33.5 to 36, shown in {\bf Figures 8-10}. The bit-reversible Milne trajectory guarantees that the trajectory is
reversed backward in time to the very last bit. Without this precaution a double-precision Runge-Kutta
trajectory will reverse only for a time of 2 and a quadruple precision for a time of 4.

The vectors corresponding to the forward and backward Lyapunov exponents for the interval
$33.5 \leq t \leq 36.0$ at the beginning of the collisional process can be visualized as in
{\bf Figures 8 and 9}, where the particles making above average contributions to the instability
are shown as filled circles. Each top+bottom pair of configurations in these {\bf Figures} is identical to
the very last bit.  But the Lyapunov exponents and their characteristic vectors differ.
In the forward compressive time direction localized important particles are reminiscent of shockwaves.
There is a more diffuse unstable region in the backward expansive direction of time, reminiscent of a
rarefaction fan.

\begin{figure}
\center
\includegraphics[width=3.5in,angle=-90,bb=87 12 524 780]{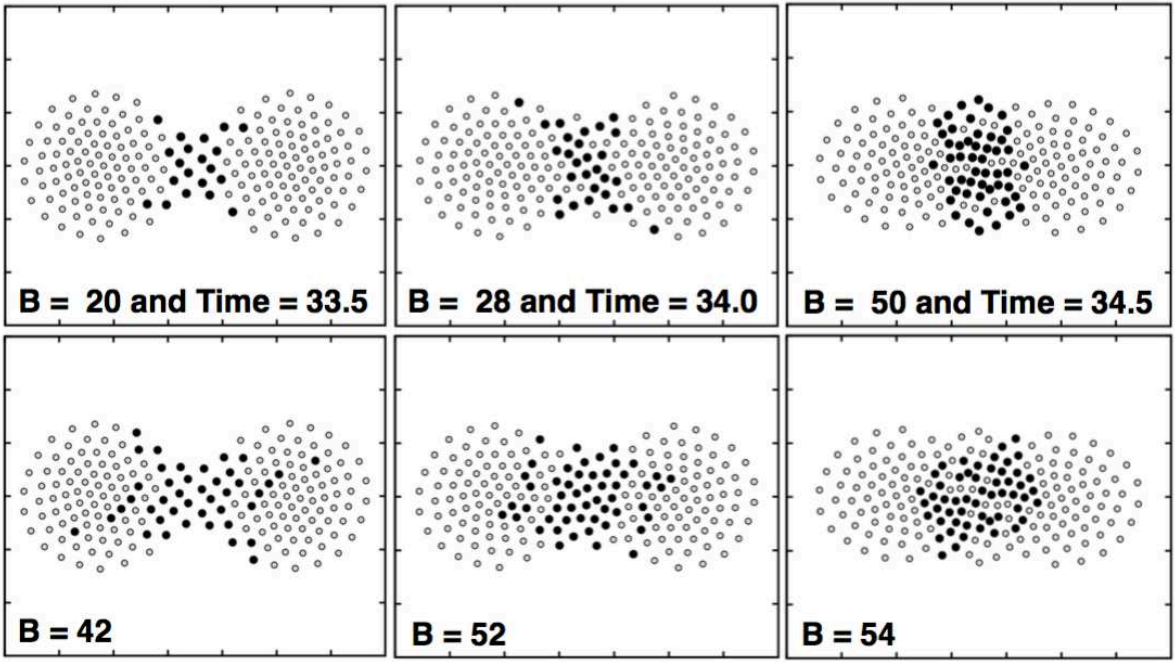}
\caption{
The local Lyapunov vector, for three configurations going forward (above) and backward (below) in
time from an original state with center-to-center distance 10 and energy -1705.  The Lyapunov data were
collected from a simulation going forward in time for 50,000 steps, with a reversal of all velocities
at time 50 followed by an evolution back to the original configuration.  The Milne algorithm made it
possible to achieve perfect fourth-order accuracy.  Notice also that the coordinates and momenta were
symmetrized at every timestep in order to assure perfect inversion symmetry of the configuration as
well as the offset vector $(\delta_q,\delta_p)$. For a time interval of 1 or 2 after reversal the local
Lyapunov exponent backward in time was the negative of that forward in time. Times shown here vary from
33.5 through 34.5 with an interval of 0.5.  The number of blacked-in particles $B$ is given for each
configuration.
}
\end{figure}

\begin{figure}
\center
\includegraphics[width=3.5in,angle=-90,bb=85 8 527 783]{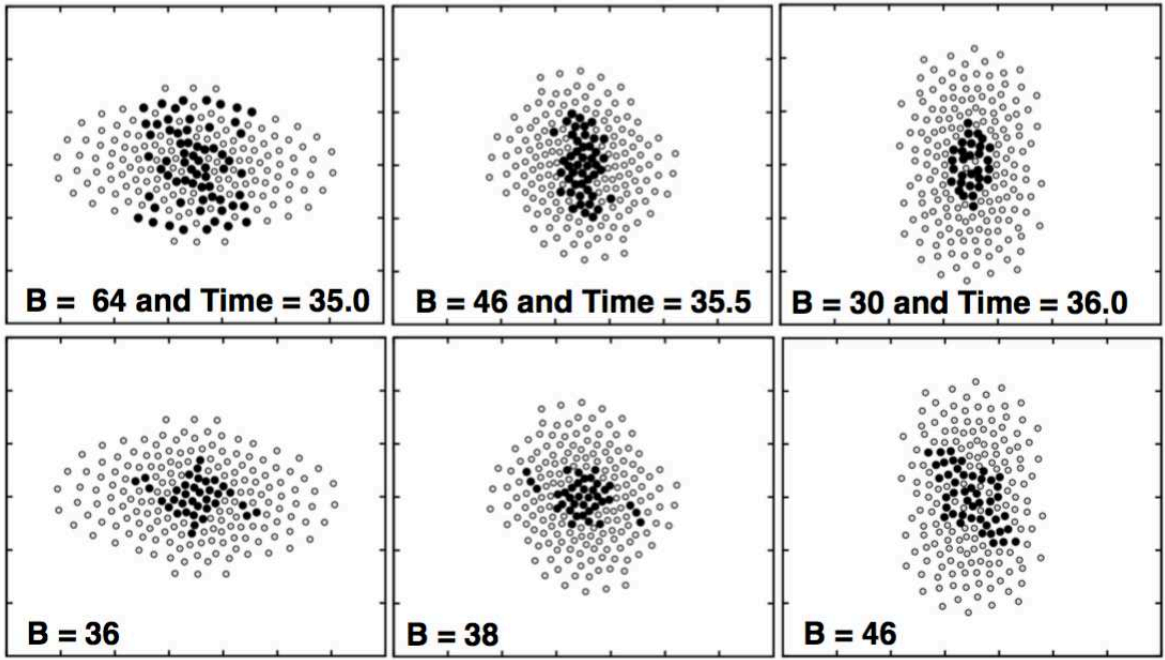}
\caption{ 
The local Lyapunov vector, for three configurations going forward (above) and backward (below) in
time from an original state with center-to-center distance 10 and energy -1705. Times shown here vary from
35 through 36 with an interval of 0.5. $B$ is the number of important particles contributing to each
of the configurations.
}
\end{figure}

\begin{figure}
\center
\includegraphics[width=3.5in,angle=-90,bb=87 8 525 781]{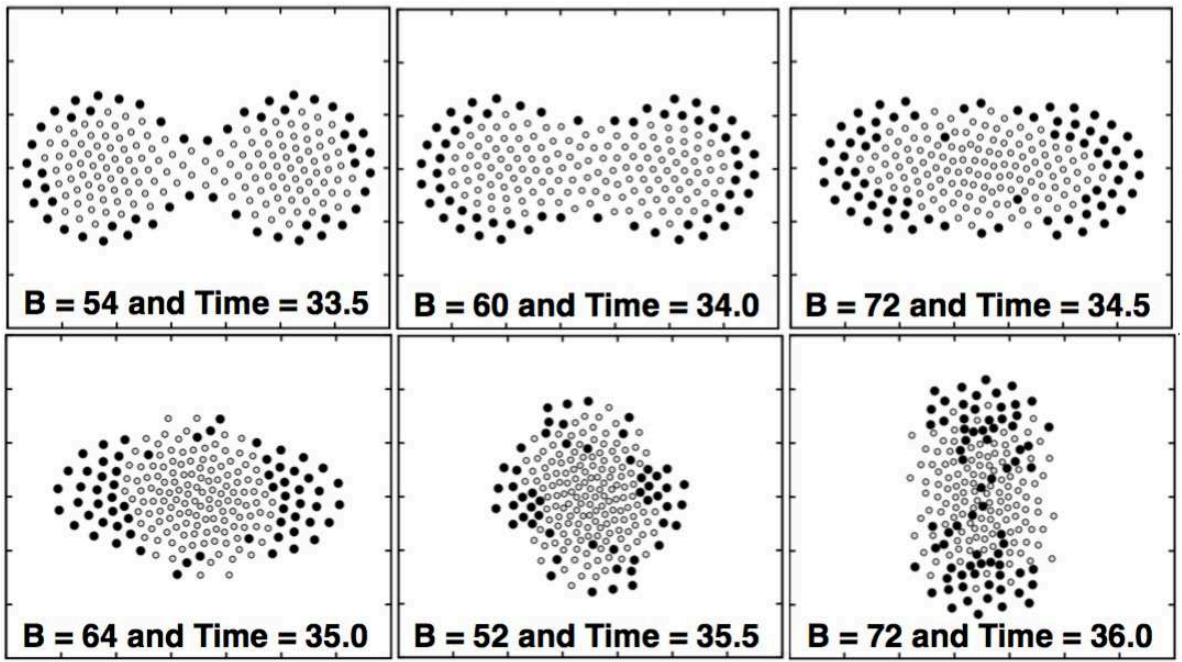}
\caption{
Particles with above-average energy are shown as filled circles.  The configurations here are identical
to those shown in {\bf Figures 8 and 9} at corresponding times.
}
\end{figure}

To add Lyapunov instability analysis to the conventional diagnostics of energy, stress, temperature, and
other functions of the particle coordinates and velocities  to the
collisional problem of Section IV we increased the offset between the two balls to 10. We symmetrized the
coordinates and velocities of the reference trajectory (calculated with Milne's algorithm) and the satellite
(calculated with Runge-Kutta integration). A timestep of 0.001 incurs errors close to double-precision
roundoff. In {\bf Figure 10} we show those particles with above-average energy for the same six times
treated in {\bf Figures 8 and 9}. In the early stages the higher-energy particles are all on the surface
and only later are affected by the high shear rate within.  The individual particle energies are exactly
the same forward and backward in time and bear no simple relationship to the forward or backward
Lyapunov vectors.

\section{Summary and Recommendation}

The bit-reversible Milne Predictor algorithm is a perfect complement to the Runge-Kutta integrator in reversibility
simulations.  Combining these two algorithms opens a highly-interesting field of study into the analysis of
irreversible proceses.  The thermodynamic state functions like temperature, energy, and pressure are all unchanged
by time reversal.  But the strain rate and heat flux both reverse along with $dt$ so that the states seen correspond
to {\it negative} values of transport coefficents.  Unlike the thermodynamic and hydrodynamic state variables local
Lyapunov exponents always react to their past rather than their future.  This turns out to be advantageous. The particles
playing a role in dynamical instability, as measured by the largest Lyapunov exponent, are markedly sensitive to the
direction of ``Time's Arrow''.  Full analyses of Lyapunov spectra for nonequilibrium processes are by now accessible
to analysis on multiple-core machines.  Evolutions forward or backward in time can be distinguished from one another
by observing the particles important to dynamic instability, ``important particles''. The present investigations are
well-suited to laptop-computer analyses and the programming of these algorithms is both challenging and educational.
We recommend the investigation of the full Lyapunov spectrum for well-chosen irreversible problems.  Such investigations
will bring further light to bear on the irreversibility hidden in Newton's reversible mechanics.

\pagebreak
 
\section*{Acknowledgments}
Over the years Harald Posch, and more recently Ken Aoki, Carl Dettmann, Puneet Patra, Clint Sprott, and Karl Travis,
have all been generous with their time and energy in helping us to our current understanding of Lyapunov instability.
Throughout this same period Daan Frenkel's innovative and generous research and fellowship have been an inspiration
and source of joy to us.

We very much appreciate Harald's comments on this manuscript and plan soon to work with him on a detailed comparison
of the Gram-Schmidt and covariant approaches to the entire Lyapunov spectrum for collisional problems of the type
introduced here.

\end{document}